# Enhanced transmission at the zero[th]-order mode of a terahertz Fabry-Perot cavity

Soumitra Hazra[1,2,†], Ran Damari[1,2], Adina Golombek[1,2], Eli Flaxer[1,3], Tal Schwartz[1,2] , and Sharly Fleischer[1,2,‡]

[1]*Raymond and Beverly Sackler Faculty of Exact Sciences, School of Chemistry, Tel Aviv University, Tel Aviv 6997801, Israel*
[2]*Tel Aviv University Center for Light-Matter Interaction, Tel Aviv 6997801, Israel*
[3]*AFEKA – Tel-Aviv Academic College of Engineering, 69107 Tel-Aviv, Israel*
Email: [†]*soumitra.hazra89@gmail.com*, [‡]*sharlyf@tauex.tau.ac.il*

**Abstract:** A planar Fabry-Perot cavity with inter-mirror spacing of $d \ll \lambda$ is explored for its "zero-order mode" terahertz transmission. The enhanced transmission observed as $d \to 0$ indicates that such cavities satisfy the resonance condition across a broad terahertz bandwidth. The experimental signatures from this elusive, "technically challenging" regime are evidenced using time-domain terahertz spectroscopy and are complemented by numerical calculations. The results raise intriguing possibilities for terahertz field modulation and pave new paths for strong coupling of multiple transition frequencies simultaneously.

**Introduction**
The Fabry-Perot interferometer, first introduced by Charles Fabry and Jean-Baptiste Alfred Perot in 1899 [1] has yielded numerous advancements and applications over more than a century [2]. From precision wavelength measurements, through atomic line shape analysis and spectral shifts, immense contributions to laser development, sensing and imaging applications, Fabry and Perot have made a huge impact on all fronts of optical and spectroscopic applications. Recently, Fabry-Perot cavities in the THz frequency range have earned much attention within the study of light-matter interaction in various systems: A two-dimensional electron gas [3], superconductors [4,5], ferromagnetic nanoparticles [6], strongly coupled organic molecules [7,8], ro-vibrational polaritons in gas phase molecules [9] and many more. The concept of strong light-matter coupling has yielded the novel , highly intriguing field of vibrational strong coupling where chemical reactions can be manipulated and controlled through zero-point energy fluctuations of the cavity's optical mode [10,11].
 A Fabry-Perot interferometer (FP) can be constructed from a pair of reflective surfaces (mirrors) positioned parallel to one another at a distance *d* apart. Light traversing the FP undergoes a series of interference events as it bounces back and forth between the mirrors. It is the multiple interferences of the light with itself as it impinges again and again on the cavity mirrors that dictate the transmission and reflection of the incoming light. Constructive interference between these partially transmitted waves at the output of the FP occurs at wavelengths satisfying the relation $d = m \cdot \frac{\lambda}{2}$ at normal incidence where $m$ is an integer. Correspondingly, the condition for resonance of an FP mode is given by $f_m = \frac{c}{2dn}m$, where $f_m$ is the resonance frequency of mode $m$, $c$ is the speed of light and $n$ the refractive index of the medium in the cavity. For many of its applications, the length of the cavity $d$ (and thus the FP mode order) is chosen based on practical considerations. For example, in a scanning FP etalon, the accuracy to which a lineshape can be determined increases with the mode number, owing to the finite accuracy with which $d$ can be set ($\Delta f = -\frac{c}{2nd^2} \cdot m \cdot \Delta d$). For strong light-matter coupling experiments however, $d$ is set such that the first FP cavity mode (m=1) frequency is resonant with the transition frequency of the material [7,10–13]. Thus, in the vast majority of FP applications, the length of the cavity is comparable to or greater than the wavelength of interest.

Previous research on optical phenomena at the deep subwavelength region have focused on harnessing the zeroth-order cavity mode to achieve perfect absorption in ultrathin film absorbers [14,15], study FP resonances at subwavelength cavities formed by sandwiched reflection gratings[16,17] and on plasmonic systems such as metallic narrow slits and sub-wavelength hole arrays [18,19]. Our current work is aimed at revealing the transmission properties of the zeroth FP cavity mode in a clear and direct way. We monitor the zero-order mode of FP in an open geometry throughout a broad range of intermirror distances, down to $d \ll \lambda$. This region possesses technical challenges in the visible or near IR wavelengths as it requires subwavelength precision ($\ll 1\mu m$) over the inter-mirror spacing. In the terahertz frequency region (THz), characterized by significantly longer wavelengths (1 $THz$ corresponds to $300 \mu m$ in wavelength), this requirement is conveniently met. In addition, the THz frequency region enables direct monitoring of the field, $E(t)$, (rather than the intensity, $I(t)$) provided by the electro-optic sampling technique [20,21]. We utilize time-domain terahertz spectroscopy (TDTS) where a single-cycle THz field traverses the FP cavity and its transmission spectrum is monitored for varying inter-mirror distances. Our results demonstrate enhanced transmission through of the FP cavity as the inter-mirror distance $d$ approaches 0.

**Experimental**

Time domain THz spectroscopy (TDTS) measurements were performed in a homebuilt TDTS setup, pumped by a femtosecond oscillator (Mai-Tai by Spectra-Physics, λ=800nm, 100fs duration). We utilize a pair of photoconductive antennas (PCA) for THz generation and detection [22] with usable bandwidth of $0.15 - 2.2\ THz$. The THz field is routed through a 4-f optical setup constructed from off-axis parabolic reflectors. The variable-length optical cavity used in this work consists of a movable mirror and a fixed mirror, as in ref. [7]. The schematic sketch of the time-domain THz spectroscopy setup is shown in Figure 1.

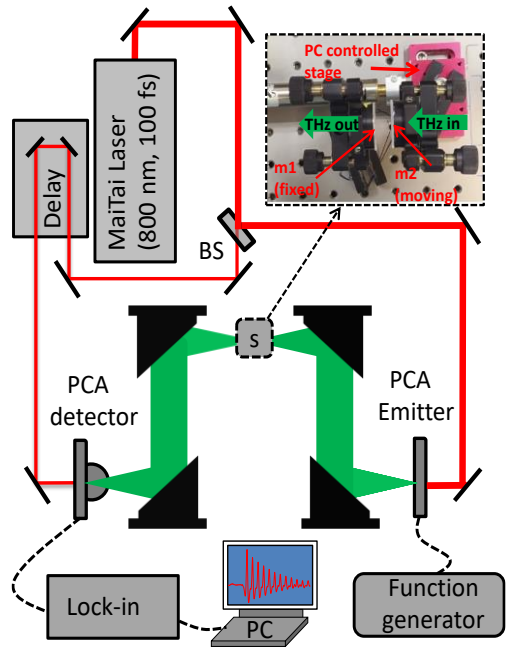

Figure 1. A scheme of the home-built time-domain THz spectrometer setup. The inset shows a photograph of the FP cavity, which consists of a fixed mirror (m1) and a moveable mirror (m2).

A green diode laser is used to set the mirrors parallel to each other by overlapping the multiple reflections from the mirrors at the far field. The cavity mirrors were fabricated by sputtering $\sim 5 nm$ of Au on a 1mm thick fused quartz substrate. While the homogeneity and smoothness of our home-sputtered thin Au films may be compromised and its morphology governed by island or percolated state, our measurements do not show any spectral responses in the THz reflection (see section S6 in the Supporting Information ). The roughness profile as characterized from the AFM scans is found to be $\sim 0.6 nm$ (RMS), i.e. negligibly small with respect to the THz wavelengths used in this work (see section S7 in the Supporting Information). We conclude that our mirrors effectively operate as a smooth and homogeneous layer for our usable THz wavelengths and in what follows are characterized by their reflectivity quantified by our THz-TDS.

**Results**

We conducted a series of TDTS transmission measurements of the FP cavity with varying inter-mirror distances and obtained a set of THz transients shown in Fig.2(a). The incidence time of the THz field was set to $t = 0$ in all of the traces of Fig.2(a).

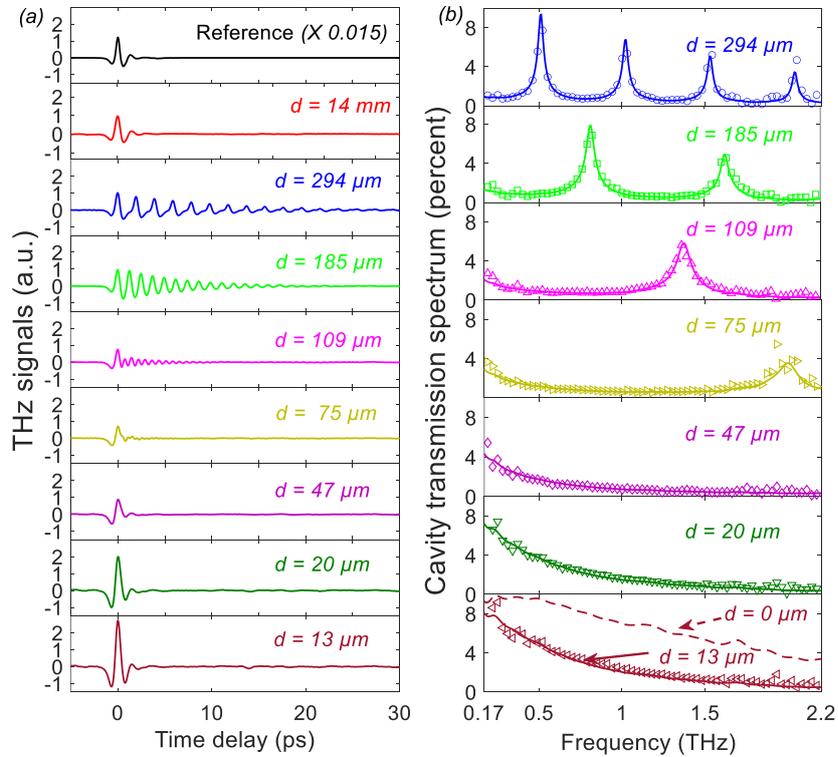

Figure 2. (a) THz transients transmitted through the FP cavity at varying cavity lengths. The incident THz field is measured without the cavity and is downscaled by a factor of 0.015 and marked 'reference' in the top pane. (b) THz transmittance spectrum of the FP cavities shown in (a). The open symbols show the experimental data. The solid lines are theoretical fits (see text).

The top panel of Fig.2a depicts the incident THz field and serves as a reference for all following transmission spectra. The red curve in Fig.2a shows the THz field transmitted through the FP cavity with $d = 14 mm$ and replicates the incident field spectrum only attenuated by $\sim 80$ fold. Note that the round-trip time of the $d = 14 mm$ cavity is $\sim 93 ps$ - far beyond the $30 ps$ span of the measurement, hence it manifests the incident field transmitted through two separated mirrors. For cavity lengths $d = 294, 185, 109 \mu m$ the transmitted field exhibits multiple reflections within the time span of the measurement. For $d = 294 \mu m$ we find 4 resonant modes at $f_1 = 0.51\, THz, f_2 = 1.02\, THz, f_3 = 1.53\, THz$ and $f_4 = 2.04\, THz$ that satisfy the relation $f_m = \frac{c}{2d} m$. These manifest as a beat signal that follows the incident THz field (blue curve in Fig.2a). As the cavity length decreases to $185\mu m$ and $109\mu m$, the fundamental resonance frequency increases, leaving two resonant cavity modes and then a single mode, respectively, within the usable THz bandwidth (see Fig.2b).

Note that the peak amplitude of the transmitted field (the very first signal peak at t=0) remains fairly constant in the range $d = 14mm \rightarrow 185\mu m$. As the cavity length is further decreased, the incident field amplitude seems to decrease in the range of $109\mu m - 75\mu m$, after which it gradually increases at $d < 75\mu m$ and the ensuing oscillatory signal is practically annihilated. We refer to this region as the zero-cavity mode ($m = 0$), where the inter-mirror distances are $d < \lambda/2$ and the first FP mode frequency is well beyond our usable bandwidth. The enhanced THz transmission is quantified by the ratio of the transient THz amplitudes for $d = 13\mu m$ and $d = 14mm$ and yields a ~2.7-fold increase. In what follows, we focus on and analyze this observation.

Figure 3 depicts the spectral amplitude transmittance of the two cavity mirrors $m_1$ and $m_2$ referenced against air (red and light green) and against the fused quartz substrate (blue and black), both required for the fitting process discussed later on. The amplitude transmittance through the FP cavity with $d = 14mm$ and $d = 13\mu m$ are depicted by the purple and green curves respectively. Each mirror alone (~6nm gold deposited on a fused quartz substrate) transmits ~13% of the THz amplitude across the usable bandwidth as can be seen from the black and blue curves in Fig.3.

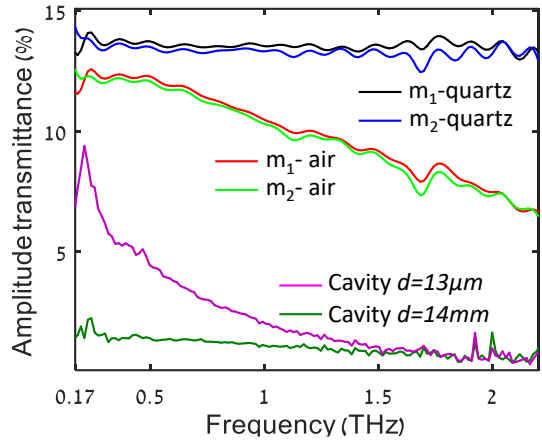

Figure 3: Frequency-resolved THz transmittance for mirror 1 and mirror 2 quantified with respect to free space (red and pale green, respectively) and with respect to the fused quartz substrate (black and blue, respectively). Transmission spectrum of the FP cavity for d=14mm (green) and d=13µm (purple).

The expected transmission through both mirrors is therefore ~1.7% ($d = 14mm$, green curve in Fig.3). For the FP cavity with $d \sim 13\mu m$ however, the amplitude transmittance reaches $> 9\%$ at the low THz frequency region $(0.17 - 0.2\ THz)$ as depicted by the purple curve, namely 5 fold larger with respect to the 'long' $(d \gg \lambda)$ inter-mirror length. The gradual decrease in transmission with frequency is clearly observed from the purple curve, and will be addressed in the remainder of the paper.

**Numerical calculations**

To calculate the spectra of Fig.2b (solid lines overlaid on the experimental data) we used the analytic transfer function (TF) of the FP cavity for monochromatic light [23,24]:

$$T_{FP}(\omega) = \left|\frac{t_1 \cdot t_2 \cdot exp(-i\varphi(\omega))}{1-r_{m1} \cdot r_{m2} \cdot exp(-2i\varphi(\omega))}\right| \quad (1)$$

$t_1$ and $t_2$ are the amplitude transmittance of the two cavity mirrors with respect to air as measured by our TDTS setup and shown in Fig.3: $t_{1,2} = E_{mirror1,2}(\omega)/E_{air}(\omega)$

The FP resonance factor is given by $1/(1 - r_{m1}(\omega) \cdot r_{m2}(\omega) \cdot \exp(-2i\varphi(\omega)))$ where $\varphi(\omega) = \omega n d/c$ is the phase accumulated through one round trip in the cavity. $n$ is the refractive index of the medium inside the cavity (air in this work) and c is the speed of light in vacuum.

The condition for FP resonance requires that the phase accumulated through one round trip is $2\pi \cdot m$ (where $m$ is an integer). For non-absorbing mirrors, this condition is met at cavity lengths of $d = m\frac{\lambda}{2}$. Metallic mirrors however, may inflict an additional phase shift upon reflection due to their absorption-loss and manifest as complex valued refraction index and corresponding complex reflection coefficient [25]. Using the complex refractive index of Au thin films of thickness 1.5nm and 8nm reported previously[26] we calculated the mirrors' reflection phase shift and found them to vary by up to 3° from $\pi$. Furthermore, from the analysis of our experimental data we obtained an upper-limit of $\frac{\pi}{50}$ to this reflection phase. We therefore conclude that the additional reflection phase in our experiment is negligibly small and restrict our analysis and discussion to real valued reflection coefficients in the remaining of this paper. Detailed discussion of the above phase shift are found in sections S4 and S5 of the Supporting Information.

The amplitude reflectance of the mirrors is obtained by fitting eq.1 to the experimental results of several FP cavities with varying lengths ranging from 305μm to 93μm (see Fig. S2(a) in the Supporting Information section S2). This range was selected as it supports multiple resonance modes within our usable THz bandwidth. The cavity length (d) used in the fitting process was calculated from the observed resonance peaks that obey: $f_m = \frac{c}{2d}m$. For simplicity we associate a single value for the reflectivity of our home-made Au sputtered mirrors [27] through our usable THz bandwidth and obtain $r_{m1} \cdot r_{m2} = 0.84 \pm 0.01$. The latter is the mean reflectivity product of the two mirrors obtained from 23 FP cavities with different lengths. The frequency-dependent amplitude reflection coefficient ($r_m$) measured for the mirrors exhibits a nearly flat response within our usable THz band, as illustrated in Fig. S6(c) in the Supporting Information section. The solid curves in Fig.2b for d ≤ 75μm were calculated using eq.1 with the extracted transmission ($t_1, t_2$) and reflection $r_{m1} \cdot r_{m2}$ noted above. The minimal cavity length we could achieve was d = 13μm due to the imperfect planarity and flatness of the mirror substrates. Note that the experimental transmission at the low THz frequency (~0.17THz) tends to coincide with the theoretically calculated transmission for d = 0 as depicted by the dashed red curve at the lower panel of Fig.2b.

**Discussion**

To explain the observed enhancement in transmission as $d \to 0$ let us begin by considering the phase accumulated by a field that satisfies the resonance condition of the FP cavity, $\lambda_m = \frac{2d}{m}$. The first resonant FP mode (with m=1) is $\lambda = 2d$, namely a resonant wavelength of the FP that propagates an exact distance of one wavelength as it traverses back and forth between the cavity mirrors (one round-trip). Considering the optical phase, a resonant field accumulates a phase of $2\pi + \pi$ (or $2\pi \cdot m + \pi$ for $m \geq 2$) throughout each round-trip, where the additional $\pi - phase$ is independent of $d$ and results purely from the Fresnel reflection off the mirror. This condition sets a constant $\pi$-phase difference between the incoming field and the inner cavity field (that propagated an integer number of round-trips in the cavity) as they overlap at the first cavity mirror. Owing to the $\pi$ phase difference between the two fields at the mirror surface, they impart counteracting forces on the mirror electrons and diminish their field-induced oscillation amplitudes. The latter effectively reduces the reflectivity of the mirror experienced by the incoming field and enhances its penetration into the cavity. This mechanistic description underlies the enhanced transmission of the FP at resonance. Using the above phase considerations, the transmission of the zero-cavity mode ($m = 0$) is readily explained: for a *d=0* cavity, the first term in the phase accumulation ($2\pi \cdot m$) vanishes for all frequency components, leaving only the (frequency-independent) π-phase shift. Correspondingly, the transmission is enhanced at the entire usable bandwidth and manifested in the enhanced THz field amplitude observed in Fig.2. Imperfections in the parallelism and flatness of the mirrors results in a minimal cavity length of $d = 13\ \mu m$ as noted above. The finite cavity length manifests as decreased transmission with frequency as observed by the purple curve in Fig.3.

To analyze the transmission of short length cavities we return to eq. (1):
$t_{FP}(\omega) = \frac{t \cdot exp(-i\varphi(\omega))}{1-r \cdot exp(-2i\varphi(\omega))}$ where $t \equiv t_1 \cdot t_2$ , $r \equiv r_{m1} \cdot r_{m2}$ and $\varphi(\omega) = \omega n d/c$

For $d = 0$ one obtains $t_{FP}(\omega) = \frac{t}{1-r}$

For $d \ll \lambda$: $exp(-i\varphi(\omega)) \cong 1 - i\varphi(\omega)$ and $exp(-2i\varphi(\omega)) \cong 1 - 2i\varphi(\omega)$

Rearranging eq.(1) with the above yields: $t_{FP}(\omega) = \frac{t}{1-r} \cdot \frac{1}{1+\frac{i\omega n d}{c}\frac{1+r}{1-r}}$ (2)

For a given reflectivity $r$ and cavity length $d$, the denominator of the second term increases with $\omega$ and the transmission decreases correspondingly. Thus, the transmission at the low THz region (~0.17 $THz$) is minimally affected by the finite length of the cavity and is found in good agreement with the calculated transmission for $d = 0$. Moreover, eq.(2) shows that as the reflectivity of the mirrors increases, so does the 'sensitivity' of the cavity to $d$ since the phase factor, $\frac{i\omega n d}{c}$ is factored by $\frac{1+r}{1-r}$ as can be seen from Fig.4a.

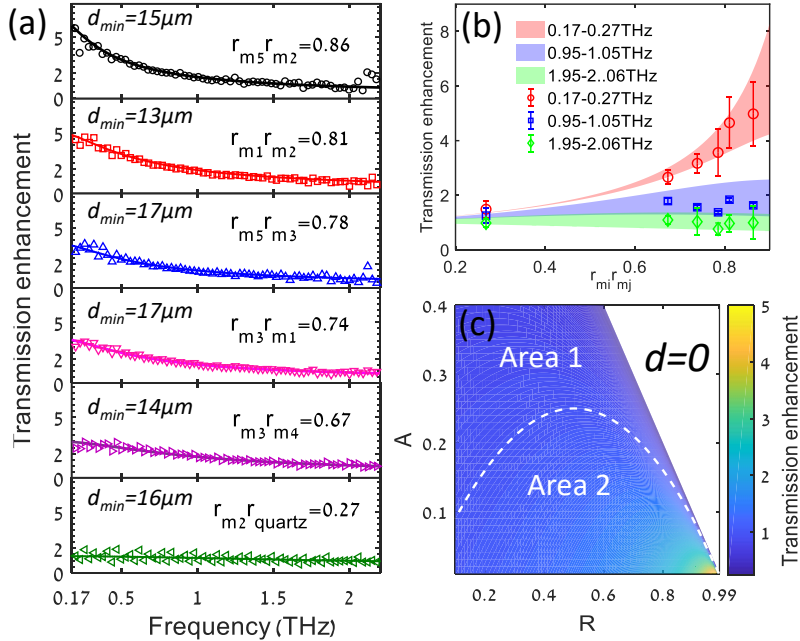

Figure 4:(a) Transmission enhancement of several cavities composed of different mirrors. The enhancement is quantified as the transmission ratio between d→0 (the minimal d obtained is noted in the figure) and d=14mm. (b) Transmission enhancement vs. the reflectivity product of the two mirrors for selected frequency bands. Open symbols depict the experimental data and pale background areas are simulated for the range of 10μm < d < 20μm (c) Theoretical enhancement map as a function of mirror reflectivity (R) and amplitude loss (A) for d=0. The enhanced transmission is quantified with respect to the transmission of a single mirror (see text). The dashed white line shows the equator between area 1 and area 2 where the cavity transmission is lower/higher than the transmission of a single mirror respectively.

Figure 4(a) depicts the enhancement factor obtained for several mirror pairs differing in Au thickness. We have fabricated mirrors 1-5 on fused quartz substrates with measured transmittance values across our usable bandwidth (0.17THz – 2.2THz marked respectively): $m_1$(13.5%-13.8%), $m_2$(13%-13.8%), $m_3$(28%-30%), $m_4$(33%-23%), and $m_5$(8.1%-8.3%). We included an untreated fused quartz substrate as a low reflectivity mirror $m_{quartz}$ (88% − 55%).
(For raw transmittance data of the mirrors see section S1 in the Supporting Information).
By pairing the different mirrors from the above cohort we constructed 6 different FP cavities ($m_1$-$m_2$, $m_3$-$m_1$, $m_3$-$m_4$, $m_5$-$m_2$, $m_5$-$m_3$ and $m_2$-$m_{quartz}$) and extracted their reflectivity product, $r_{mi} \cdot r_{mj}$, by fitting to eq. 1, (using the same protocol of Fig.2, see Fig S2(b) in the Supporting Information). The minimal cavity lengths obtained were in the range of $13\mu m < d_{min} < 17\mu m$. Fig.4a depicts the enhanced transmission for the different cavities across our THz bandwidth. The enhancement in transmission was experimentally obtained by $\frac{T_{m_i m_j}(d \to 0)}{T_{m_i m_j}(d \to \infty)}$ (open symbols) and calculated by $\left|\frac{e^{-i\varphi}}{1-r_{m_i}r_{m_j}e^{-2i\varphi}}\right|$ (solid curves), where $m_i$ and $m_j$ are indexes for the different mirrors. The results in Fig.4a show that the enhancement factor increases with the reflectivity of the FP mirrors. Figure 4b depicts the measured enhancement for three representative frequency bands (0.17-0.27THz, 0.95-1.05THz and 1.95-2.05THz). The filled areas (color coded in red, blue and green, respectively) show the simulated enhancement for cavity lengths $10\mu m < d < 20\mu m$. The low frequency band ~$0.2THz$ best represents the enhancement expected for $d \to 0$ owing to its 'long' wavelength. The enhancement obtained for higher frequency bands (~1THz and ~$2THz$ depicted by the blue and green curves, respectively) is strongly compromised due to the deviation of $d$ from 0.

Inspired by the enhanced transmission observed as $d \to 0$ we set out to explore the breadth of the effect. Different from the above, where the enhancement is quantified by the ratio of FP with $d \to 0$ and of the FP with (effective) $d \to \infty$, here we focus on a more refined quantity - the transmission ratio between the FP with d=0 and a single mirror. Using the results of Fig. 3 we quantify this ratio as ~0.8 for the low frequency component (~$0.17THz$), namely the experimental transmission of the FP remains lower than that of a single mirror. Fig.4c shows a theoretical prediction for this ratio for varying mirror parameters. For simplicity, we consider a FP cavity comprised of two identical mirrors with amplitude reflectance $r_1 = r_2 = r$, amplitude transmission $t_1 = t_2 = t$ and attenuation loss (A). These three parameters are bound by the conservation relation: $t = \sqrt{1-r^2-A}$. Thus, for a given $A$ and $r$, we extract the mirror transmission $t$ and plot the ratio of the FP cavity transmission (T) and that of the mirror, given by $\frac{T}{t} = \left|\frac{t}{1-r^2}\right|$. For high reflectivity (r=0.98) and low attenuation loss (A=0.01) we predict a ~5 fold enhancement ($\frac{T_{d=0}}{t}$). This simplified calculation predicts that for mirror parameters within area II (marked in Fig.4c) where the attenuation loss is sufficiently low and the mirror reflectance sufficiently high – one can increase the transmission through a single mirror by a few-fold by merely placing another identical mirror next to the first. Note that the above calculation assumes a zero distance between the FP mirrors ($d = 0$). Since the practical flatness of any mirror is finite, $d = 0$ is beyond experimental reach. However, tuning the cavity length to $d \sim 1\mu m$ or even smaller is possible [28]. Sufficiently short cavity length is expected to manifest as enhanced transmission without significant compromise to the spectral content (see Supporting Information section S3).

**Conclusions**

The zero-order mode of a Fabry-Perot cavity manifests enhanced transmission across a broad THz bandwidth. This enhanced transmission suggests that the zero-order mode satisfies the resonance condition for the entire spectrum. The latter may offer uniquely intriguing possibilities in the realm of strong-coupling phenomena, e.g. the simultaneous coupling of multiple transitions of different frequencies within the same cavity. Technical challenges in accessing the zero-order regime are significantly eased by the long wavelength of THz field and enable direct experimental exploration of this typically elusive region.

**Funding**. The authors acknowledge the support of the Israel Science Foundation (926/18, 1856/22, 1435/19, 1993/13), the Wolfson family foundation (PR/ec/20419) and the PAZI foundation.

**Disclosures.** The authors declare no conflicts of interest.

**Data availability.** Data underlying the results presented in this paper are not publicly available at this time but may be obtained from the authors upon reasonable request.

# Supporting Information

## Section S1: Amplitude transmittance of the mirrors and reference THz spectrum

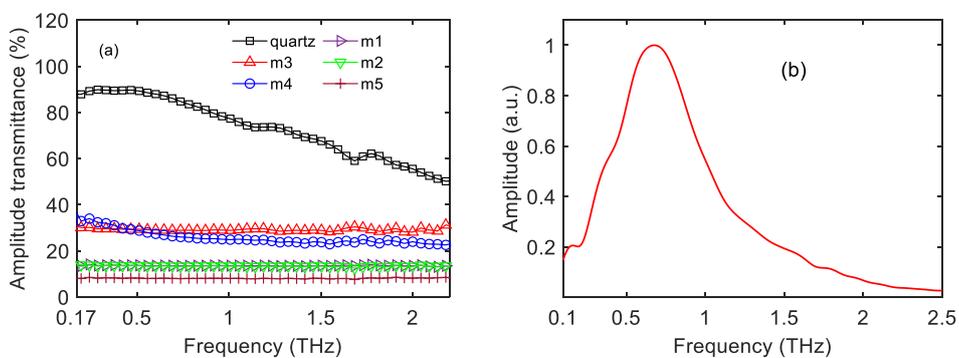

**Figure S1:** (a) The THz amplitude transmittance through different mirrors (m1, m2, m3, m4, m5) and fused quartz. (b) Reference THz spectrum measured in air.

# Sections S2: Extraction of amplitude reflectance via fitting to cavity transmission

Fig. S2(a) depicts the FP cavity transmission at varying lengths (*d*) in the range 305μm→109μm. By fitting to eq.1 in the main text file, we extract the reflectivity product of mirrors $m_1$ and $m_2$. Figure S2(b) represents the transmission of the FP cavity for the cavity lengths 595μm, 448μm, 248μm, 555 μm, 262μm and 320μm for 6 mirror pairs: $m_5$-$m_2$, $m_1$-$m_2$, $m_5$-$m_3$, $m_3$-$m_1$, $m_3$-$m_4$, and $m_2$-quartz respectively. We extracted the reflectivity products of the above mirror pairs ($r_{mi}r_{mj}$) by fitting the cavity transmission to eq.1.

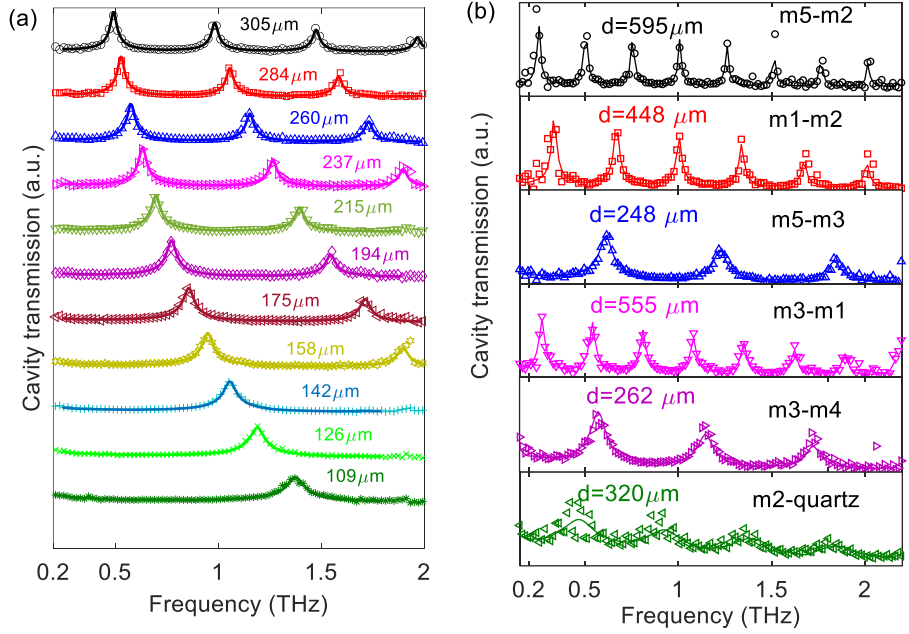

**Figure S2:** **(a)** Cavity amplitude transmittance fitted with the FP transfer function (eq. (1) in the main text) for different cavity lengths of FP cavity (m1-m2) **(b)** The cavity transmission for six mirror pairs: m5-m2, m1-m2, m5-m3, m3-m1, m3-m4, and m2-quartz with the cavity lengths 595μm, 448μm, 248μm, 555 μm, 262μm and 320μm respectively. The experimental data is given by the open symbols and the theoretical fit to eq.1 is depicted by the solid lines.

## Sections S3: Transmission enhancement of FP cavity at $d \neq 0 \mu m$ compared to a single mirror transmission

Transmission amplitude of the FP calculated for several cavity lengths using Eq. (1). The cavity is composed of two identical mirrors with parameters R=0.85, A=0.1. The black line depicts the amplitude transmission of a single mirror and the colored lines show the FP transmission at cavity lengths ($0 \mu m \leq d \leq 2 \ \mu m$). As noted in the text, we predict enhanced transmission for finite cavity lengths with respect to the transmission of a single mirror. Reduction in the enhancement factor is observed at $d > 0 \ \mu m$, nevertheless, for $d = 0.5 \mu m$ the modulation in the THz spectral content is practically negligible (blue curve).

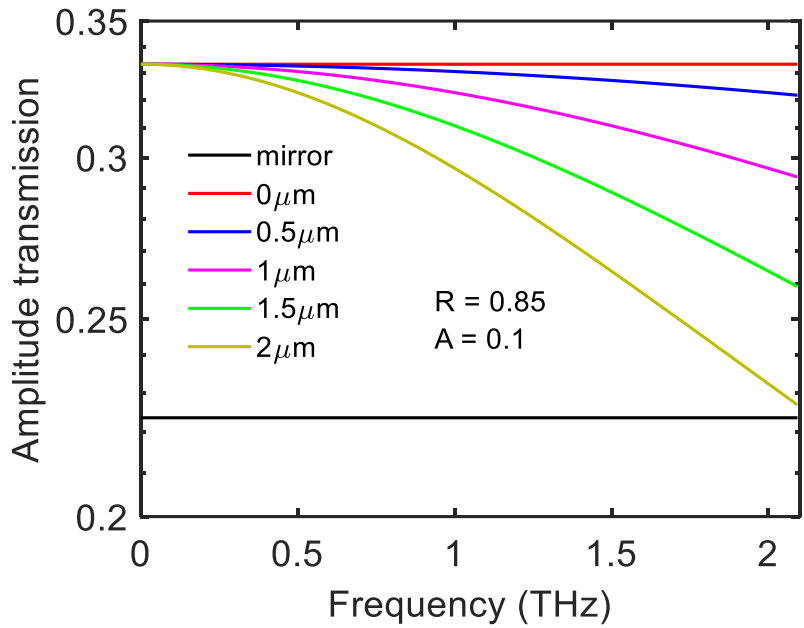

**Figure S3:** Amplitude transmission of the FP cavity calculated for several cavity lengths ($0 \mu m \leq d \leq 2 \ \mu m$) using the FP transfer function (Eq. (1) in the main text file). The FP cavity consists of identical mirrors with parameters R=0.85 and A=0.1. The transmission amplitude of the mirror is shown by the black solid line.

**Table S1**. Experimentally obtained parameter r$_{m1}$r$_{m2}$ after fitting the FP cavity transfer function (Eqn. (1)) to the cavity transmittance at different cavity lengths for the single mirror pair m$_1$-m$_2$.

| Cavity length (μm) | Fitted parameters ($r_{m1}r_{m2}$) | |
|---|---|---|
| 305 | 0.8294 | |
| 294 | 0.8358 | |
| 284 | 0.8349 | |
| 272 | 0.8286 | |
| 260 | 0.8300 | |
| 249 | 0.8340 | |
| 237 | 0.8409 | **Mean ($r_{m1}r_{m2}$): 0.83826** |
| 226 | 0.8308 | |
| 215 | 0.8346 | **Standard deviation: 0.00615** |
| 204 | 0.8337 | |
| 194 | 0.8432 | |
| 185 | 0.8361 | |
| 175 | 0.8416 | |
| 166 | 0.8491 | |
| 158 | 0.8489 | |
| 149 | 0.8493 | |
| 142 | 0.8426 | |
| 133 | 0.8407 | |
| 126 | 0.8361 | |
| 118 | 0.8361 | |
| 109 | 0.8355 | |
| 101 | 0.843 | |
| 93 | 0.8409 | |

## Sections S4: Phase shift at the reflection of air-mirror interface

In general, the phase shift of a field reflected from the air-metal interface may differ from $\pi$, owing to the absorption loss in metallic mirrors. In this section and in section S5 we show that for our experimental parameters, the distortion of the phase from $\pi$ is practically negligible. This validates our choice to treat the reflection as a real valued coefficient as done throughout the main text. In order to calculate the phase of the reflected field we used the complex refractive index reported previously for a 1.5nm thick Au thin film and for 8nm thick Au film (ref.[26]). We have digitized the real (Fig. 3c in ref. 26) and imaginary part (Fig.4c in ref. 26) to obtain the frequency dependent complex refractive index of the two samples.

The complex reflection coefficient is calculated using the expression outlined in ref. [29]:

$$\tilde{r} = \frac{r_{01} + r_{12} e^{-i\frac{4\pi n_1 d}{\lambda}}}{1 + r_{01} r_{12} e^{-i\frac{4\pi n_1 d}{\lambda}}} \quad (s1)$$

Where $r_{01}$, $r_{12}$ is the Fresnel reflection coefficient of the air-film and film-quartz interfaces respecitvely, given by $r_{ij} = \frac{n_j - n_i}{n_j - n_i}$, with $n_0 = 1$ (air) is the refractive index of air, $n_1 = n_{Au} - ik_{Au}$ (thin Au film) and $n_2 = 1.955$ is the refractive index of the fused quartz substrate (Fig. S5 in the supplementary information of ref [30]. $d$ is the Au film thickness. Figure S4 shows the phase calculated from the reflection coefficient of expression (s1) for the case of 1.5nm Au film (blue curve) and 8nm Au film (red curve).

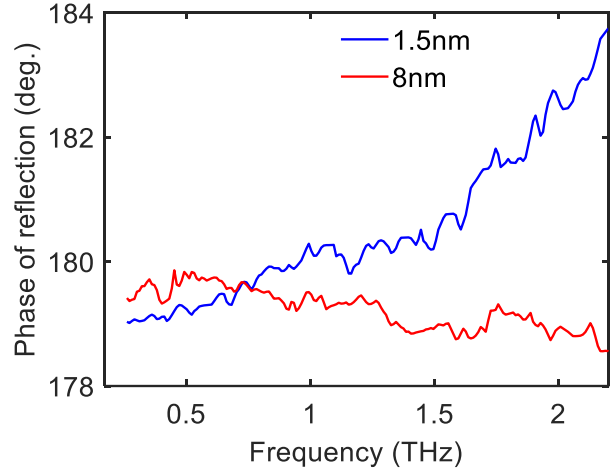

Figure S4 : the reflection phase calculated for our Au-on-fused quartz substrates of 5nm thickness using the complex refractive index extracted from ref.1 for 1.5nm Au film (Blue curve) and for 8nm Au film (Red curve).

Thus for our 5nm Au film (as quantified in section S7), the phase shift of our sample is bound by the blue and red curves, namely in the range of [178°, 183°]. We conclude that the maximal phase shift of our reflected field is < 4° for the entire usable bandwidth and ≤ 1° for frequencies lower than 1.5THz, validating its negligible distortion from $\pi$ and our choice to attribute a real-valued reflection coefficient throughout the text.

# Sections S5: Phase distortion obtained from cavity transmission measurements

In order to further validate the negligibility of the phase shift due to reflection from the air-mirror interface, we returned to our experimental measurements. Here we rely on our ability to monitor multiple FP orders at a fixed cavity length supported by our several octaves spanning THz spectrum. For example, see the blue curve in Fig.2b (with $d = 294\ \mu m$) where we resolve the first 4 resonant frequencies of orders m=1,2,3,4. Those are just an harmonic series of the lowest frequency, and as long as the additional phase concerned here is 0 (i.e. the reflected phase is purely $\pi$), the resonance frequencies are equally spaced from one another. Once we include an additional phase in the calculation (a phase that emanates from absorption loss of the Au film), the cavity modes gradually shift and become non-equally spaced anymore.

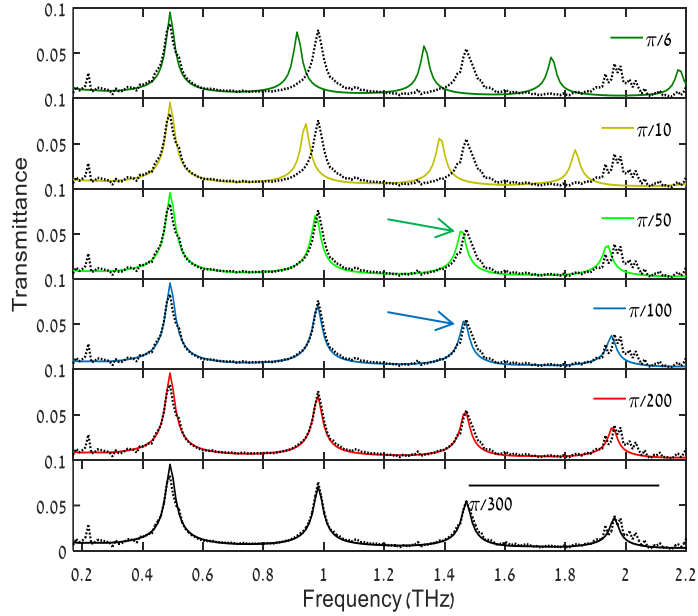

Figure S5: Comparison of the experimental data (black dashed curves) with calculated FP transmission where we 'inject' varying additional phases ($\psi$) noted in each panel.

The calculation is performed using:

$$T_{FP}(\omega) = \left|\frac{t_1 \cdot t_2 \cdot \exp(-i\varphi(\omega))}{1-|r_{m1}|\cdot|r_{m2}|\cdot\exp(-2i(\varphi(\omega)+\psi))}\right|$$

By comparing the experimental results with the calculated transmission curves we extract an upper limit for $\psi$, relying on our spectral experimental accuracy to detect a phase shift as small as $\psi = \pi/50$ as readily observed by difference at ~1.5THz (marked by an arrow). For $\psi = \pi/100$ we can hardly resolve the difference between the experimental and calculated peak at ~1.5THz, while the difference at ~2THz is still reasonably observed (despite the noise). We therefore conclude that the additional phase shift imparted by our mirrors is smaller than $\pi/50$, providing clear experimental justification for neglecting this phase and correspondingly treating the reflection coefficients as real-valued throughout the main text file.

## Section S6: THz reflection measurement

In order to further verify the quality of our home-made mirrors, we've measured their frequency dependent reflectivity using the THz-TDS setup in reflection configuration shown in Fig.S6(a). Here the angle of incidence is dictated by the off-axis parabolic reflectors at $15^0$ to the sample normal. The reflectance of our mirrors is quantified with respect to that of a commercially available gold mirror (PF10-03-M01,Thorlab).

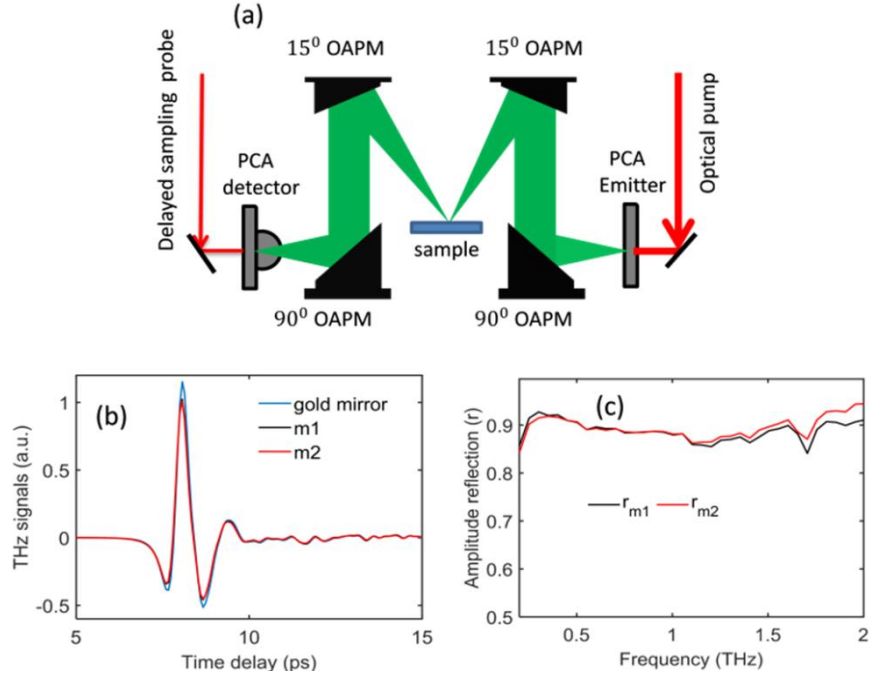

Figure S6: (a) schematic sketch of the THz-TDS reflection setup. (b) The time-domain THz field reflected off the mirrors (m1 and m2) and off the reference reflector (gold mirror, PF10-03-M01 from Thorlabs). (c) Frequency resolved amplitude reflection of mirrors $m_1$ and $m_2$.

The schematic of the THz setup in the reflection geometry is shown in Fig. S6(a). The THz field is routed by four off-axis parabolic mirrors (OAPM). The first $90^0$ OAPM collimates the THz beam and the second $15^0$ OAPM focuses the THz beam onto the sample surface. The THz field reflected from the sample is collected and collimated by the third $15^0$ OAPM and finally focused by the last $90^0$ OAPM onto the PCA detector.

As can be readily observed in time domains signal of Fig.S6(b) the THz field reflected from the mirror is attenuated to about 90% from of that of the reference, with similar temporal shape. The frequency-dependent reflectance coefficient ($r_{mi}$) is obtained by $|r_{mi}| = \left|\frac{E_{mi}}{E_{gold-mirror}}\right|$, is shown in Fig. S6(c) with no apparent modulations across the usable THz spectrum, i.e. showing a 'flat' response as expected from homogeneous metallic surface.

## Section S7: Atomic Force Microscopy image of representative mirror

The morphology of the mirrors surface was characterized by atomic force microscope (AFM) operating at room temperature in a contact mode and the images were analyzed using the Gwyddion software. The film thickness was measured by the height difference across a designated step created by scratching the film surface with a sharp razor and scanning across the step as shown in Fig S7 (a). The white line illustrates the scanning segment across the designated step and the corresponding height profile is shown in Fig. S7 (b). The film height is 5nm. In order to quantify the surface morphology we have scanned an area of 0.5µm x 0.5µm shown in Fig. S7 (c) which exhibits a uniform and continuous surface with surface roughness less than 1nm. The roughness profile is shown in Fig. S7 (d).

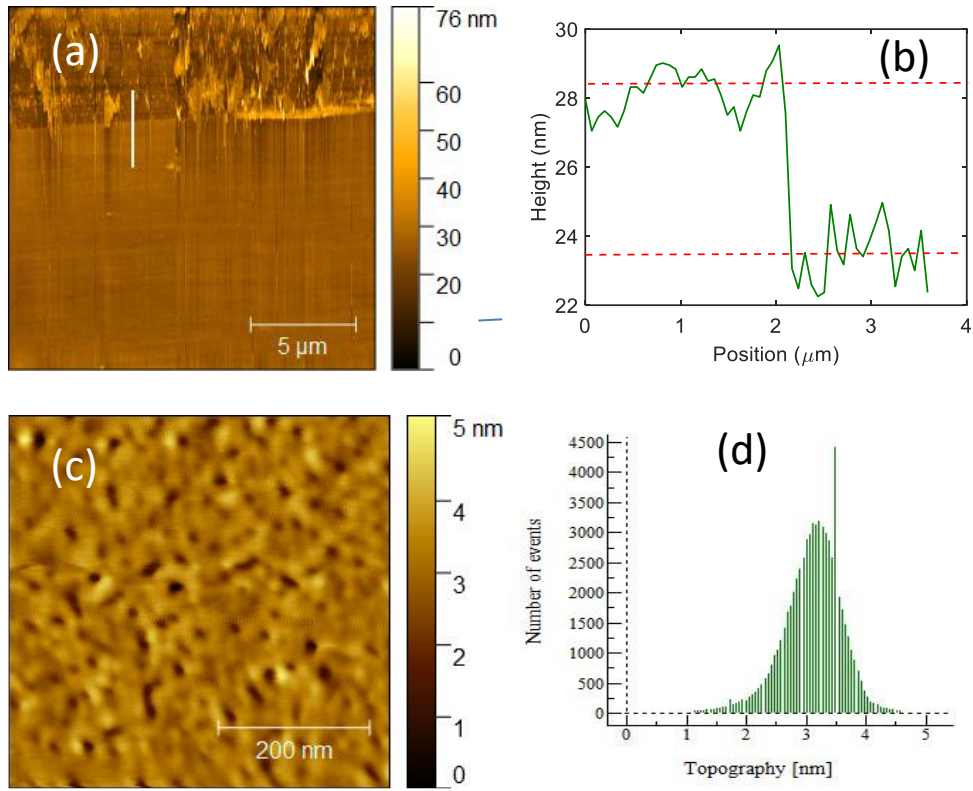

Figure S7: (a) AFM height segment (b) height profile (c) AFM image showing the surface morphology of the mirror. (d) Roughness profile with mean square roughness of 0.6nm.